\documentclass[pra,showpacs,letterpaper,showpacs,superscriptaddress]{revtex4}
\usepackage{graphicx,amsmath,amssymb,amsfonts,latexsym,color,dcolumn,bm}

\begin{document}

\newcommand{\la}{\langle}
\newcommand{\ra}{\rangle}
\newcommand{\be}{\begin{equation}}
\newcommand{\ee}{\end{equation}}
\newcommand{\bea}{\begin{eqnarray}}
\newcommand{\eea}{\end{eqnarray}}
\newcommand{\om}{\omega}
\newcommand{\Om}{\Omega}
\newcommand{\omo}{\omega_0}
\newcommand{\pa}{\partial}
\newcommand{\eps}{\epsilon}
\newcommand{\bE}{{\bf E}}
\newcommand{\bH}{{\bf H}}
\newcommand{\cE}{{\cal E}}
\newcommand{\br}{{\bf r}}
\newcommand{\bd}{{\bf d}}
\newcommand{\bk}{{\bf k}}
\newcommand{\bR}{{\bf R}}
\newcommand{\sig}{\sigma}
\newcommand{\sigz}{\sigma_z}
\newcommand{\sigd}{\sigma^{\dag}}
\newcommand{\bek}{{\bf e}_{{\bf k}\lambda}}
\newcommand{\subk}{_{{\bf k}\lambda}}
\newcommand{\rP}{{\rm P}}

\title{Van der Waals Interactions in a Magneto-Dielectric Medium}

\author{S. Spagnolo}
\affiliation{CNISM and Dipartimento di Scienze Fisiche ed Astronomiche, Universita degli Studi di Palermo, Via Archirafi 36,
I-90123 Palermo, Italy}
\affiliation{Theoretical Division, MS B213, Los Alamos National Laboratory, Los Alamos, New Mexico 87545, USA}

\author{D.A.R. Dalvit}
\affiliation{Theoretical Division, MS B213, Los Alamos National Laboratory, Los Alamos, New Mexico 87545, USA}

\author{P.W. Milonni}
\affiliation{104 Sierra Vista Dr., Los Alamos, New Mexico 87544, USA}

\date{\today}

\begin{abstract}

The van der Waals interaction between two ground-state atoms is calculated for two electrically or magnetically polarizable particles embedded in a dispersive magneto-dielectric medium. Unlike previous calculations which infer
the atom-atom interaction from the dilute-medium limit of the macroscopic, many-body van der Waals interaction, the interaction is calculated directly for the system of two atoms in a magneto-dielectric medium.
Two approaches are presented, the first based on the quantized electromagnetic field in a dispersive medium
without absorption and the second on Green functions that allow for absorption. We show that the correct
van der Waals interactions are obtained regardless of whether absorption in the host medium is taken into account.
\end{abstract}


\maketitle

%%%%%%%%%%%%%%%%%%%%%%%%%%%%%%%%%%%%%%%%%%%%%%%%%%%

\section{Introduction}

The van der Waals interaction between ground-state atoms in vacuum is often regarded as a consequence of the vacuum fluctuations of the electromagnetic field. It is well known that the interaction between two macroscopic, uncharged dielectric bodies cannot be obtained by pairwise addition of this interatomic van der Waals interaction except in the dilute-medium limit. In this limit Lifshitz, for instance, obtained the retarded and unretarded pairwise van der Waals interaction between electrically polarizable atoms from a more general expression for the interaction energy of two dielectric half-spaces \cite{Lifshitz56}. The van der Waals interaction obtained in this way is identical to that obtained more directly by Casimir and Polder \cite{CP48} for the system of two atoms in vacuum.

The continuing interest in Casimir effects and the related effects of vacuum field fluctuations in the case of
dielectric media has stimulated some interest in the van der Waals interactions of atoms embedded in magneto-dielectric media.
Recent work in this area \cite{TomasPRA05}-\cite{Welsch2atoms},
takes essentially the original approach of Abrikosov {\it et al.}
\cite{Abrikosovbook} using Green functions for radiation in an {\it absorbing} medium and taking the dilute-medium limit to
infer the atom-atom interaction. In this paper we obtain the atom-atom van der Waals interaction {\it directly} rather
than from a dilute-medium limit of an interaction between macroscopic bodies, and we show that this interaction can
be correctly obtained {\it without explicit account of absorption in the host medium}.

We consider both electrically polarizable and magnetically polarizable atoms and present two derivations. The first,
presented in Section \ref{sec2}, is based on the (electric or magnetic) dipole-dipole interaction induced in the two atoms by the ``vacuum" field in the magneto-dielectric medium. For this purpose we employ simple expressions for the quantized electromagnetic field in a dispersive magneto-dielectric medium in which absorption is ignored \cite{pwmetc}. The second, the subject of Section \ref{sec3}, employs Green functions and takes explicit account of absorption \cite{quantabsorption}. The fact that the same results are obtained regardless of whether absorption is accounted for appears to us to be of some interest, and  physical reasons for it are given in Section \ref{sec4}, which also includes some further discussion and a brief summary of our results.

%%%%%%%%%%%%%%%%%%%%%%%%%%%%%%%%%%%%%%%%%%%%%%%%%%

\section{\label{sec2} Van der Waals Interactions from Quantized Field without Absorption}

We consider two identical atoms in a homogeneous and isotropic magneto-dielectric medium which we regard as a continuum with real electric permittivity $\epsilon(\om)$ and magnetic permeability $\mu(\om)$, and therefore
real refractive index $n(\om)$. The calculation of the van der Waals interactions in this section will be based on the following expressions for the electric and magnetic fields in the non-absorbing magneto-dielectric \cite{Milonni2003}:
\be
\bE(\br,t)=i\sum\subk\left( \frac{ 2\pi\hbar\om_k\mu_k }{ n_k\gamma_kV}\right)^{1/2}[a\subk(t)e^{i\bk\cdot\br}
-a\subk^{\dag}(t)e^{-i\bk\cdot\br}]\bek ,
\label{electricfield}
\ee
\be
\bH(\br,t)=i\sum\subk\left( \frac{2\pi\hbar c^2 }{ \om_kn_k\gamma_k\mu_kV}\right)^{1/2}[a\subk(t)e^{i\bk\cdot\br}
-a\subk^{\dag}(t)e^{-i\bk\cdot\br}] \bk\times\bek .
\label{magneticfield}
\ee
We employ a standard notation in which $a\subk(t),a\subk^{\dag}(t)$ are Heisenberg-picture photon annihilation and
creation operators for the plane-wave mode with wave vector $\bk$ [$|{\bk}|=k=n_k\om_k/c$, $n_k=n(\om_k)$] and (linear)
polarization unit vector $\bek$, [$\bk\cdot\bek=0$, $\lambda=1,2$]. The refractive index is $n_k=(\eps_k\mu_k)^{1/2}$, where
$\eps_k$ and $\mu_k$ are the electric permittivity and magnetic permeability, respectively, at frequency $\om_k$.
The group index is $\gamma_k\equiv n_k+\om_kdn_k/d\om_k$, and $V$ is the quantization volume for the box
normalization of the plane-wave modes. The Hamiltonian is
\be
H = H_A+H_B+H_F - \bd_A(t) \cdot \bE(\br_A,t) - \bd_B(t)  \cdot \bE(\br_B,t)
- {\bf m}_A(t) \cdot \bH(\br_A,t) - {\bf m}_B(t) \cdot \bH(\br_B,t) ,
\label{Hamiltonian}
\ee
where $H_A$, $H_B$, and $H_F$ are the Hamiltonian operators for atom A, atom B, and the electromagnetic
field, respectively. The electric dipole moment operators are denoted by  $\bd(t)$ and the magnetic
dipole moment operators by ${\bf m}(t)$.

%%%%%%%%%%%%%%%%%%%%%%%%%%%%%%%%%%%%%%%%%%%%%%%%%%%%%

\subsection{Van der Waals interaction between electrically polarizable particles}

We first consider the van der Waals interaction between two
electrically polarizable atoms separated by a distance $R$ in the
magneto-dielectric medium. The approach we will take follows
closely that used by various authors for the derivation of the van
der Waals interaction in the case in which the particles are in
vacuum ($n=1$) \cite{vdwpapers}. The basic idea of this method is
that quantum vacuum fluctuations of the electromagnetic field in
the medium induce in the atoms fluctuating electric and magnetic
dipoles that interact with each other. The electric dipole moment
induced in an atom at $\br$ by an electric field is $\bd(t) =
\alpha_e(t) \bE(\br,t)$, where $\alpha_e$ is the (real) electric
polarizability of the atom. (Since it is only the real part of the
polarizability that determines shifts in energy levels, we can
assume without loss of generality throughout this paper that the
polarizabilities of the guest atoms are real.) The dipole
interaction energy between the two induced, fluctuating electric
dipoles is then
 \be W_{ee}(R)  = \sum_{\bk \lambda} \alpha^A_e(\om)
\alpha^B_e(\om) \langle \bE_i(\br_A,\bk \lambda) \bE_j(\br_B,\bk
\lambda) \rangle  V^{ee}_{ij}(\om, {\bf R}) , 
\label{ee_interacion}
\ee
where ${\bf R}=\br_B-\br_A$, $R=|{\bf R}|$. The two-point
vacuum electric-field correlation function, summed over
polarization states, follows easily from (\ref{electricfield}):
\be \sum_{\lambda} \langle \bE_i(\br_A,\bk \lambda)
\bE_j(\br_B,\bk \lambda) \rangle = \frac{2 \pi \hbar \omega_k
\mu_k}{n_k \gamma_k V} (\delta_{ij} - \hat{{\bf k}}_i \hat{{\bf
k}}_j) e^{-i {\bf k} \cdot {\bf R}} , 
\label{EEcorrelator} \ee
with $\hat{{\bf k}}_i = {\bf k}_i / k$. The interaction potential
$V^{ee}_{ij}(\om, {\bf R})$ between two oscillating electric dipoles
embedded in the magneto-dielectric is calculated as follows. From
the Hamiltonian (\ref{Hamiltonian}) and the Heisenberg equations
of motion for the annihilation and creation operators, one obtains
a formal expression for the electric field generated by the
electric dipole B at the position of the electric dipole A: \be
{\bE}_e(\br_A,t) = \frac{i}{\pi c^3} \int_0^{\infty} d\om n(\om)
\mu(\om) \om^3 \left[ {\bf a}  \frac{\sin k R}{k R} - {\bf b}
\left(  \frac{\sin k R}{k^3 R^3} - \frac{\cos k R}{k^2 R^2}
\right) \right] \int_0^t dt' p_B(t') e^{i\om(t'-t)} + h.c. ,
\label{electricdipole} \ee where ${\bf a} =
\bd_B-(\bd_B\cdot\hat{{\bf R}})\hat{{\bf R}}$, ${\bf
b}=\bd_B-3(\bd_B\cdot\hat{{\bf R}})\hat{{\bf R}}$, and $\hat{{\bf
R}} = {\bf R} / R$. We have used the notation $\bd(t)=\bd p(t)$,
with $\bd$ a unit vector specifying the direction of the electric
dipole moment. We are interested in the case of two electric
dipoles, both oscillating at frequency $\om'$: $p_B(t)=C_B
e^{-i\om't} + C_B^*e^{i\om't}$ and $p_A(t)=C_A e^{-i\om't} +
C_A^*e^{i\om't}$, where $C_A, C_B$ are arbitrary constants. We
define the interaction $V^{ee}(\bR) = -\bd_Ap_A(t)\cdot
{\bE}_e(\br_A,t)$. After performing the time integration for times
$t\gg 1/\om'$, and taking the time average of the resulting
expression, we get \be V^{ee}(\bR) = - \frac{2}{\pi c^3}{\rm
Re}\int_0^{\infty}d\om n(\om)\mu(\om)\om^3 \left[a \frac{\sin kR}{
kR}-b\left(\frac{\sin kR }{ k^3R^3}- \frac{\cos kR }{
k^2R^2}\right)\right] \left[ \frac{C_A^*C_B }{\om-\om'-i\eta}+
\frac{C_AC_B^* }{\om+\om'-i\eta}\right] , \ee where as usual $\eta
\rightarrow 0^+$. Here $a=\bd_A\cdot\bd_B-(\bd_A\cdot\hat{{\bf
R}})(\bd_B\cdot\hat{{\bf R}})$ and
$b=\bd_A\cdot\bd_B-3(\bd_A\cdot\hat{{\bf R}})(\bd_B\cdot\hat{{\bf
R}})$. We can also define the interaction as
$V^{ee}(\bR)=-\bd_Bp_B(t)\cdot {\bE}_e(\br_B,t)$, which amounts to
interchanging A and B above. This implies that we can take
$C_A^*C_B=C_B^*C_A$. Note also that the time average
$\overline{p_A(t)p_B(t)}=C_AC_B^*+C_A^*C_B$, so that we can write
\be V^{ee}(\bR)=- \frac{1}{\pi c^3}\overline{p_Ap_B} \;  {\rm Re}
\int_0^{\infty}d\om n(\om)\mu(\om)\om^3\left[a \frac{\sin kR }{
kR}-b\left( \frac{\sin kR }{ k^3R^3}- \frac{\cos kR }{
k^2R^2}\right)\right]\left( \frac{1}{\om-\om'-i\eta}+ \frac{1 }{
\om+\om'-i\eta}\right) . \label{Vee} \ee Next we use the fact that
$n$ and $\mu$ (or actually their real parts which are implicit
here) are even functions of $\om$ to rewrite this as \be
V^{ee}(\bR) =  - \frac{1}{\pi c^3}\overline{p_Ap_B}
\int_{-\infty}^{\infty} d\om n(\om)\mu(\om)\om^3\left[a \frac{\sin kR }{ kR}-b\left( \frac{\sin kR}{ k^3R^3}-\frac{\cos kR }{ k^2R^2}\right)\right] \left(\frac{1 }{ \om-\om'-i\eta}+ \frac{1}{\om-\om'+i\eta}\right) . \nonumber \\
\ee
Performing the trivial contour integrations, writing $\om$ instead of $\om'$ for the dipole frequencies, and
using again $k=n(\om)\om/c$, we obtain the electric dipole-dipole interaction tensor needed in
Eq. (\ref{ee_interacion}):
\be
V^{ee}_{ij}(\om,{\bf R}) = \frac{1}{\epsilon_k} \frac{1}{R^3}
[ (\delta_{ij} -  3 \hat{{\bf R}}_i \hat{{\bf R}}_j)  (\cos kR+ kR \sin kR) - (\delta_{ij} - \hat{{\bf R}}_i \hat{{\bf R}}_j)  k^2R^2\cos kR]  .
\label{electric_dipole_tensor}
\ee

Using Eqs. (\ref{EEcorrelator}) and (\ref{electric_dipole_tensor}), and passing to the continuum limit
$\sum_{{\bf k}}\longrightarrow (V/8\pi^3)\int_0^{\infty}dk k^2\int
d\Omega_{\mathbf{k}}=(V/8\pi^3c^3)\int_0^{\infty}d\omega
\gamma_{\omega}n^2_{\omega}\omega^2\int d\Omega_{\mathbf{k}}$,
we obtain the van der Waals interaction energy between electrically polarizable particles:
\be
W_{ee}(R) = -\frac{\hbar}{\pi
c^3}\frac{1}{R^3}\int_0^{\infty}d\omega
\alpha^A_e(\omega) \alpha^B_e(\omega)\frac{\omega^3 \mu^2(\omega)}{n(\omega)} \;
\left[
kR\sin 2kR+2\cos 2kR-5\frac{\sin 2kR}{kR}-6\frac{\cos 2kR}{k^2R^2}+3\frac{\sin 2kR}{k^3R^3}
\right] .
\ee
The integration path can be rotated using the fact that there are no there are no poles in the upper
half of the complex plane. We obtain finally
\be
W_{ee}(R)=-\frac{\hbar}{16\pi R^6}\int_0^{\infty}du
\alpha^A_e(iu) \alpha^B_e(iu)\frac{1}{\epsilon^2(iu)} F\left[ \frac{2n(iu) uR}{c} \right] e^{-2n(iu)uR/c} ,
\label{result1}
\ee
where $F(x)=x^4+4x^3+20x^2+48x+48$. Recall that along the imaginary frequency
axis the electric permittivity and refractive index are real and positive. The electric-electric van der Waals force that results from Eq.(\ref{result1}) is therefore always attractive, regardless of the frequency dependence of $\eps(\om)$ and $n(\om)$ .

Our calculation based on the quantized fields (\ref{electricfield}) and (\ref{magneticfield})
gives a van der Waals interaction (\ref{result1}) in full agreement with that obtained recently by
Toma$\check{s}$ \cite{TomasJPA06}, for instance. The same is true for the other van der Waals interactions
we calculate in this paper. The main point of this section is to show that correct results for
van der Waals interactions involving ground-state atoms in dispersive media can be obtained straightforwardly, without
having to go to a dilute-medium limit of an interaction between macroscopic bodies, and without having to
introduce complexities arising from absorption. We discuss this further in Section \ref{sec4}.

%%%%%%%%%%%%%%%%%%%%%%%%%%%%%%%%%%%%%%%%%%%%%%%%%%%%%%%%%%%%%%

\subsection{Van der Waals interaction between magnetically polarizable particles}

We next use the same approach to calculate the van der Waals
interaction between magnetic dipoles induced in the atoms by
fluctuations of the zero-point magnetic field. For this the
relation between an induced magnetic dipole moment at position
${\bf r}$ and the magnetic field is ${\bf m}(t) = \alpha_m(t)
\bH({\bf r},t)$, where $\alpha_m$ is the (real) magnetic
polarizability of the atom. The dipole interaction between the two
induced, fluctuating magnetic dipoles is 
\be 
W_{mm}(R)
=\sum_{\bf{k} \lambda} \alpha^A_m(\om) \alpha^B_m(\om) \langle
\bH_i({\bf r}_A, {\bf k} \lambda) \bH_j({\bf r}_B, {\bf k}
\lambda) \rangle V^{mm}_{ij}(\om, {\bf R}) , \label{magint} 
\ee
where the two-point vacuum magnetic field correlation function
summed over polarizations is found from (\ref{magneticfield}) to
be 
\be 
\sum_{\lambda}\langle \bH_i(\br_A,\bk \lambda)
\bH_j(\br_B,\bk \lambda) \rangle = \frac{2 \pi \hbar n_k
\omega_k}{\mu_k\gamma_k V } (\delta_{ij} - \hat{{\bf k}}_i
\hat{{\bf k}}_j) e^{-i {\bf k} \cdot {\bf R}} .
\label{HHcorrelator} 
\ee 
One can derive the magnetic dipole-dipole
interaction tensor following steps similar to those above for the
electric dipole-dipole interaction: 
\be 
V^{mm}_{ij}(\om,{\bf R})=
\frac{1}{\mu_{k}}\frac{1}{R^3} [  (\delta_{ij} - 3 \hat{{\bf R}}_i
\hat{{\bf R}}_j) (\cos k R + k R \sin k R)- (\delta_{ij} -
\hat{{\bf R}}_i \hat{{\bf R}}_j)  k^2 R^2 \cos k R]  ,
\label{magnetic_dipole_tensor} 
\ee 
which differs from
(\ref{electric_dipole_tensor}) simply by the replacement of
$\epsilon_k$ by $\mu_k$. The details of the evaluation of
(\ref{magint}) are essentially the same as for the electric van
der Waals interaction and lead straightforwardly to the expression
\be 
W_{mm}(R)=- \frac{\hbar}{16\pi R^6} \int_0^{\infty}du
\alpha^A_m(iu) \alpha^B_m(iu) \frac{1}{\mu^2(iu)} F\left[
\frac{2n(iu) uR}{c} \right] e^{-2n(iu)uR/c} . 
\label{result2} 
\ee
Recall that along the imaginary frequency axis the magnetic
permeability is real and positive. The magnetic-magnetic van der Waals force that results from
Eq.(\ref{result2}) is always attractive, regardless of the frequency dependence of $\mu(\om)$ and $n(\om)$.

%%%%%%%%%%%%%%%%%%%%%%%%%%%%%%%%%%%%%%%%%%%%%%%%%%%%%

\subsection{Van der Waals interaction between an electrically polarizable particle and a magnetically polarizable particle}
In calculating $W_{ee}(R)$ and $W_{mm}(R)$ it has not been necessary to account for the fact that the field operators
in (\ref{ee_interacion}) and (\ref{magint}) do not commute. Because of this noncommutativity, it is more appropriate
to write $W_{ee}(R)$, for instance, in the symmetrized form
\bea
W_{ee}(R) &=& \frac{1}{ 2}\sum_{\bk \lambda} \alpha^A_e(\om)
\alpha^B_e(\om)\left[ \langle \bE_i(\br_A,\bk \lambda)
\bE_j(\br_B,\bk\lambda) \rangle + \langle \bE_j(\br_B,\bk \lambda)\bE_i(\br_A,\bk\lambda) \rangle\right]
 V^{ee}_{ij}(\om, {\bf R}) \nonumber \\
&=&{\rm Re}\sum_{\bk \lambda} \alpha^A_e(\om)
\alpha^B_e(\om)\langle \bE_i(\br_A,\bk \lambda) \bE_j(\br_B,\bk\lambda) \rangle  V^{ee}_{ij}(\om, {\bf R}).
\label{ee_interacion2}
\eea
The forms (\ref{EEcorrelator}) and (\ref{electric_dipole_tensor}), however, show that symmetrization
is actually not required because the summation over ${\bf k}$ does not require us to distinguish between
$\langle \bE_i(\br_A,\bk \lambda) \bE_j(\br_B,\bk\lambda) \rangle$ and $\langle \bE_j(\br_B,\bk \lambda)
\bE_i(\br_A,\bk\lambda) \rangle$.

The situation in the case of the van der Waals interaction between an electrically polarizable particle and
a magnetically polarizable particle, however, is different because the electric-magnetic correlation function summed over polarization states, 
\be \sum_{\lambda} \langle \bE_i({\bf r}_A,{\bf k}\lambda)
 \bH_j ({\bf r}_B,{\bf k}\lambda) \rangle = \frac{2 \pi \hbar
\omega_k }{\gamma_k V } \epsilon_{ijl} {\hat{\bf k}}_l e^{-i {\bf
k} \cdot {\bf R}} , 
\label{neweq1}
\ee
is not purely real when summed over ${\bf k}$. ($\epsilon_{ijl}$ is the Levi-Civita tensor.) Moreover the
interaction tensor in this case, which we calculate to be
\be
V_{ij}^{em}(\om,{\bf R})= \frac{\om^3}{ c^3}n^2(\om)\eps_{ijp} \hat{{\bf R}}_p 
\left[ \frac{\sin kR}{ k^2R^2}-\frac{\cos kR}{ kR}\right]
\label{neweq2}
\ee
in a manner directly analogous to the electric-electric and magnetic-magnetic tensors, is antisymmetric.

Let ${\bf E}_m({\bf r}_A,t)$ be the electric field (operator) at ${\bf r}_A$ due to a magnetic dipole at ${\bf r}_B$.
We write the interaction between the fluctuating electric and magnetic dipole moments in the symmetrized form
\be 
W_{em}(R)=-{\rm Re}\sum_{{\bf k}\lambda}\alpha^A_e(\om)\langle {\bf E}_{i}^{(+)}({\bf r}_A,{\bf k}\lambda) {\bf E}_{mi}^{(-)}({\bf r}_A,
{\bf k}\lambda)\rangle,
\label{neweq3}
\ee
where ${\bf E}^{(+)}({\br}_A,t)$ is the positive-frequency (photon annihilation) part of the source-free (``vacuum")
electric field operator at ${\bf r}_A$, and ${\bf E}^{(-)}_m({\br}_A,t)$ is the negative-frequency
(photon creation) part of the electric field produced by the magnetic dipole moment at ${\bf r}_B$.
This electric field  is induced by the source-free magnetic field ${\bf H}({\bf r}_B,t)$, so that the evaluation of (\ref{neweq3})
involves the electric-magnetic correlation function (\ref{neweq1}). The calculation is essentially just the
same as that presented by Farina {\it et al} \cite{farina} for the case where the two particles are in free space, except of course that in our case the refractive index $n(\om)$ appears:
\be
W_{em}(R)=\frac{\hbar }{ 4\pi c^2 R^4} \int_0^{\infty} du u^2 \alpha^A_e(iu) \alpha^B_m(iu) 
G\left[ \frac{2n(iu)uR}{ c}\right]e^{-2n(iu)uR/c},
\label{neweq4}
\ee
where $G(x) = (x+2)^2$. $n(iu)$ is real and positive, so that $W_{em}(R)$ is always repulsive, regardless of the frequency dependence of the refractive index.

%%%%%%%%%%%%%%%%%%%%%%%%%%%%%%%%%%%%%%%%%%%%%%%%%%%%%%

\section{\label{sec3} Van der Waals Interactions from Quantized Field with Absorption}

In this section we will calculate the van der Waals interactions considered in the previous section for two atoms
embedded in a magneto-dielectric, {\it but now taking absorption in the host medium into account}.
We use the quantization procedure for the EM field in a dispersive and absorbing medium based on the Green-function formulation \cite{quantabsorption}. The dyadic Green function ${\bf G}({\bf r},{\bf r}',\omega)$ satisfies
\cite{quantabsorption}
\be
\left[  \nabla \times  \kappa({\bf r},\omega) \nabla \times - \frac{\omega^2}{c^2} \epsilon({\rm r}, \omega) \right]
{\bf G}({\bf r},{\bf r}',\omega) = \delta({\bf r},{\bf r}',\omega) ,
\label{green_function}
\ee
as well as the appropriate boundary conditions.
Here $\kappa({\bf r},\omega) = \mu^{-1}({\bf r},\omega)$. In an infinite, homogeneous material,
\be
G_{ij}({\bf r}, {\bf r}',\omega) =  \frac{\mu(\omega)}{4 \pi k^2(\omega)}
\left[
k^2(\omega) ( \delta_{ij} - \hat{{\bf R}}_i \hat{{\bf R}}_j ) - ( \delta_{ij} - 3 \hat{{\bf R}}_i \hat{{\bf R}}_j)
\left( \frac{1}{R^2} - \frac{i k(\omega)}{R} \right)
\right]
\frac{e^{i k(\omega) R}}{R} ,
\ee
where ${\bf R} = {\bf r} - {\bf r}'$, $R=| {\bf R}|$, $\hat{{\bf R}} = {\bf R}/R$, and $k(\omega) = n(\omega) \omega /c$.
The refractive index of the medium is
given by $n^2(\omega) = \epsilon(\omega) \mu(\omega) $, with
$\epsilon(\omega) = \epsilon'(\omega) + i \epsilon''(\omega)$ the complex electric permittivity
and
$\mu(\omega) = \mu'(\omega) + i \mu''(\omega)$ the complex magnetic permeability.

The quantized electric field in dispersive, absorbing media may be written in the form \cite{quantabsorption}
\be
{\bE}({\bf r},\omega) = \sum_{\lambda=e,m} \int d^3 {\bf r}' {\bf G}_{\lambda}({\bf r}, {\bf r}', \omega) \cdot
{\bf f}_{\lambda}({\bf r}', \omega) + h.c. ,
\ee
where the operators ${\bf f}_{\lambda}$ are bosonic operators satisfying the
usual commutation relations:
\be
[ f_{\lambda,i}({\bf r}, \omega), f^{\dagger}_{\lambda',j}({\bf r}', \omega') ]
= \delta_{\lambda,\lambda'} \delta_{ij} \delta(\omega-\omega')   ~~~; ~~~
[ f_{\lambda,i}({\bf r}, \omega), f_{\lambda',j}({\bf r}', \omega') ] = 0 .
\ee
These operators may be regarded as being variables of the system composed
of the EM field {\it and} the medium including the dissipative system.
The electric and magnetic dyadic Green functions are defined in terms of the full Green function as
\bea
{\bf G}_e({\bf r},{\bf r}', \omega) &=&
i \frac{\omega^2}{c^2} \sqrt{ \frac{\hbar}{\pi} {\rm Im} \epsilon({\bf r}',\omega)}
\; {\bf G}({\bf r},{\bf r}', \omega) ,  \\
{\bf G}_m({\bf r},{\bf r}', \omega) &=&
-i \frac{\omega^2}{c^2} \sqrt{- \frac{\hbar}{\pi } {\rm Im} \kappa({\bf r}',\omega)}
\; [ {\bf G}({\bf r},{\bf r}', \omega) \times \nabla_{{\bf r}'} ] .
\eea
Note that for an absorbing medium ${\rm Im} \epsilon({\bf r}, \omega) > 0$,
${\rm Im} \mu({\bf r}, \omega) > 0$, and ${\rm Im} \kappa({\bf r}, \omega) < 0$.
The quantized magnetic field, similarly, may be written as
\be
{\bH}({\bf r},\omega) = \sum_{\lambda=e,m} \frac{c}{i \omega \mu(\omega)}
\int d^3 {\bf r}' \nabla_{{\bf r}} \times  {\bf G}_{\lambda}({\bf r}, {\bf r}', \omega) \cdot
{\bf f}_{\lambda}({\bf r}', \omega) + h.c. ,
\ee
and the total Hamiltonian for the free field is
\be
\hat{H} = \sum_{\lambda=e,m} \int d^3 {\bf r} \int_0^{\infty} d\omega  \hbar \omega
{\bf f}^{\dagger}_{\lambda}({\bf r}, \omega) \cdot {\bf f}_{\lambda}({\bf r},\omega) .
\ee
We will require the following two-point vacuum field correlation functions obtained from these expressions:
\bea
\langle  \bE_{i}({\bf r},\omega) \bE_{j}^{\dagger}({\bf r}',\omega')  \rangle &=&
\frac{\hbar}{\pi} \frac{\omega^2}{c^2} \delta(\omega-\omega')
{\rm Im} [G({\bf r}, {\bf r}', \omega) ]_{ij} , \\
\langle  \bH_{i}({\bf r},\omega) \bH_{j}^{\dagger}({\bf r}',\omega')  \rangle &=&
\frac{\hbar}{\pi} \frac{1}{|\mu(\omega)|^2 } \delta(\omega-\omega')
{\rm Im}[ \nabla_{{\bf r}} \times \nabla_{{\bf r}'} \times  G({\bf r}, {\bf r}', \omega)]_{ij} , \\
\langle \bE_{i}({\bf r},\omega) \bH_{j}^{\dagger}({\bf r}',\omega') \rangle &=&
- \frac{1}{i \omega \mu^*(\omega)}
\frac{\hbar}{\pi} \frac{\omega^2}{c} \delta(\omega-\omega')
{\rm Im} [ \nabla_{{\bf r}'} \times G({\bf r}, {\bf r}', \omega) ]_{ij} .
\eea

The electric and magnetic fields at the position of the atom A are given by the sum of the ``vacuum"
contributions, $\bE_0({\bf r}_A,\omega)$ and $\bH_0({\bf r}_A,\omega)$, plus the fields
generated by the atom B which contain both electric dipole and magnetic dipole components.
An electric dipole located at position ${\bf r}_B$ generates fields at position ${\bf r}_A$ given by
\bea
\bE_e({\bf r}_A, \omega) &=&  \omega^2   \alpha^B_e(\omega)
{\bf G}({\bf r}_A,{\bf r}_B, \omega) \cdot \bE_0({\bf r}_B, \omega)  + h.c. , \\
\bH_e({\bf r}_A, \omega) &=& -i   \omega c \kappa(\omega)  \alpha^B_e(\omega)
[ \nabla_{{\bf r}_A} \times {\bf G}({\bf r}_A,{\bf r}_B, \omega) ] \cdot \bE_0({\bf r}_B, \omega) + h.c.  ,
\eea
while the fields generated by a magnetic dipole at position ${\bf r}_B$ are
\bea
\bE_m({\bf r}_A, \omega) &=& -i \omega  \kappa(\omega) \alpha^B_m(\omega)
[ \nabla_{{\bf r}_A} \times {\bf G}({\bf r}_A,{\bf r}_B, \omega) ] \cdot \bH_0({\bf r}_B, \omega) + h.c. , \\
\bH_m({\bf r}_A, \omega) &=&  c  \kappa(\omega)  k^2(\omega) \alpha^B_m(\omega)
{\bf G}({\bf r}_A,{\bf r}_B, \omega) \cdot \bH_0({\bf r}_B, \omega) + h.c.
\eea

The vacuum expectation value of the van der Waals energy may be written as the sum of three
contributions, one purely electric, one purely magnetic, and one mixed. Using the above expressions
for the two-point correlation functions of the EM field, one can easily find each of these terms. The purely electric
part stems from the ${\bf p} \cdot \bE$ interaction, and is found to be
\bea
W_{ee}(R) &=&   - \frac{1}{2} \int_0^{\infty} d\omega d\omega' \alpha^A_e(\omega) e^{i(\omega-\omega')t}
\langle  \hat{\bE}_0({\rm r}_A,\omega) \cdot  \hat{\bE}^{\dagger}_e({\bf r}_A,\omega')  \rangle + h.c. \nonumber \\
&=& - \frac{ \hbar}{\pi} \int_0^{\infty} d\omega \alpha^A_e(\omega) \alpha^B_e(\omega)
\omega^4 {\rm Re}[G({\bf r}_A,{\bf r}_B, \omega)]_{ij} {\rm Im}[G({\bf r}_A,{\bf r}_B, \omega)]_{ij} .
\eea
Comparing this expression with Eq.(\ref{ee_interacion}) we see that $\omega^2 {\rm Im}[G({\bf r}_A,{\bf r}_B, \omega)]_{ij}$ is related
to the electric dipole-dipole interaction tensor $V^{ee}_{ij}(\om, {\bf R})$, and that
$\omega^2 {\rm Re}[G({\bf r}_A,{\bf r}_B, \omega)]_{ij}$ is related to the solid-angle integration of the
two-point vacuum electric-field correlation function summed over polarization states, given in Eq.(\ref{EEcorrelator}).
After rotation in the complex plane ($\omega \rightarrow i u$), we can re-write this expression as
\be
W_{ee}(R) = - \frac{ \hbar}{2 \pi } \int_0^{\infty} du \alpha^A_e(i u) \alpha^B_e(i u)
u^4 {\rm Tr} [G({\bf r}_A,{\bf r}_B, i u)  \cdot G({\bf r}_A,{\bf r}_B, i u) ].
\label{green_electric}
\ee
The purely magnetic part, similarly, comes from the ${\bf m} \cdot \bH$ interaction:
\bea
W_{mm}(R) &=&   - \frac{1}{2} \int_0^{\infty} d\omega d\omega' \alpha^A_m(\omega) e^{i(\omega-\omega')t}
\langle \hat{\bH}_0({\rm r}_A,\omega) \cdot  \hat{\bH}^{\dagger}_m({\bf r}_A,\omega')  \rangle + h.c. \nonumber \\
&=& - \frac{ \hbar}{\pi} \int_0^{\infty} d\omega \alpha^A_m(\omega) \alpha^B_m(\omega)
\frac{c}{|\mu(\omega)|^2}  {\rm Re} [ \kappa(\omega) k^2(\omega) G({\bf r}_A,{\bf r}_B, \omega)]_{ij}
{\rm Im}[ \nabla_{{\bf r}_A} \times \nabla_{{\bf r}_B} \times G({\bf r}_A,{\bf r}_B, \omega)]_{ij}  \nonumber \\
&=& - \frac{ \hbar}{2 \pi } \int_0^{\infty} du \alpha^A_m(i u) \alpha^B_m(i u)
u^4 \frac{\epsilon^2(i u)}{\mu^2(i u)}
{\rm Tr} [G({\bf r}_A,{\bf r}_B, i u)  \cdot G({\bf r}_A,{\bf r}_B, i u) ] ,
\label{green_magnetic}
\eea
where Eq.(\ref{green_function})  and a rotation in the complex plane were
used in obtaining the last equality.
Finally, there are two electric-magnetic terms, one arising from the ${\bf p} \cdot \bE$ interaction, and one
from the ${\bf m} \cdot \bH$ interaction. They result in the mixed interaction
\bea
W_{em}(R) &=&
-\frac{1}{2} \int_0^{\infty} d\omega d\omega' e^{i (\omega-\omega')t}
[ \alpha_e^A(\omega) \langle \hat{\bE}_0({\rm r}_A,\omega) \cdot  \hat{\bE}^{\dagger}_m({\bf r}_A,\omega')  \rangle
+ \alpha_m^A(\omega) \langle \hat{\bH}_0({\rm r}_A,\omega) \cdot  \hat{\bH}^{\dagger}_e({\bf r}_A,\omega')  \rangle
+ h.c. ] \nonumber \\
&=&
\frac{ \hbar}{\pi} \int_0^{\infty} d\omega [ \alpha^A_e(\omega) \alpha^B_m(\omega) +
\alpha^A_m(\omega) \alpha^B_e(\omega) ]
\frac{\omega^2}{c}  {\rm Re}[  \kappa^2(\omega) \nabla_{{\bf r}_A} \times G({\bf r}_A,{\bf r}_B, \omega)]_{ij}
{\rm Im}[ \nabla_{{\bf r}_B} \times G({\bf r}_A,{\bf r}_B, \omega)]_{ij} \nonumber \\
&=&
\frac{ \hbar}{2 \pi } \int_0^{\infty} du [ \alpha^A_e(i u) \alpha^B_m(i u) +
\alpha^A_m(i u) \alpha^B_e(i u) ]
u^2 {\rm Tr} [ \nabla_{{\bf r}_A} \times G({\bf r}_A,{\bf r}_B, i u)
\cdot  \nabla_{{\bf r}_B} \times G({\bf r}_A,{\bf r}_B, i u)] .
\label{green_mixed}
\eea
The traces appearing in the integrands of  Eqs.(\ref{green_electric},\ref{green_magnetic},\ref{green_mixed}) can be
explicitly computed given the form of the dyadic Green function evaluated at the imaginary frequency $w=i u$.
The final result for the complete van der Waals interaction energy between two ground state atoms embedded in an
absorbing and dispersive medium is then
\bea
W(R) &=& -  \frac{\hbar}{16 \pi R^6} \int_0^{\infty}
du e^{-2 n(i u) u R/c} F\left[ \frac{2 n(i u) R}{c} \right]
\left[ \frac{\alpha^A_e(i u) \alpha^B_e(i u)}{\epsilon^2(i u)} +
\frac{\alpha^A_m(i u) \alpha^B_m(i u)}{\mu^2(i u)} \right] \\ \nonumber
&+&
\frac{\hbar}{4 \pi c^2 R^4} \int_0^{\infty}
du u^2 e^{-2 n(i u) u R/c} G\left[ \frac{2 n(i u) R}{c} \right]
\left[ \alpha^A_e(i u) \alpha^B_m(i u) + \alpha^A_m(i u) \alpha^B_e(i u) \right] ,
\eea
where again $F(z) = z^4+4 z^3+20 z^2 +48 z +48$ and $G(z)=(z+2)^2$. This is identical to the complete van der Waals
interaction obtained in Section \ref{sec2}.

%%%%%%%%%%%%%%%%%%%%%%%%%%%%%%%%%%%%%%%%%%%%%%%%%%%%%%%%

\section{\label{sec4} Discussion}

Since the van der Waals interaction between electrically polarizable particles is the most important,
a rough model for the modification of the vacuum interaction by the medium might be of interest.
Let us consider a two-level model in which the polarizabilities are
\be
\alpha_e^A(\om)=\alpha_e^B(\om)\equiv\alpha(\om)= \frac{2\om_0d^2/3\hbar }{\om_0^2-\om^2},
\label{mod1}
\ee
where $d$ and $\om_0$ are respectively the (real) transition electric dipole moment and the
transition angular frequency, and similarly
\be
n(\om)=[1+4\pi N\alpha(\om)]^{1/2},
\label{mod2}
\ee
where $N$ is the atomic density of the host medium and we take $\mu=1$. It is useful to normalize (\ref{result1}) to the
familiar, nonretarded London form of the interaction:
\be
W_L(R)=- \frac{3\hbar\om_0\alpha^2(0) }{4 R^6}=- \frac{3\hbar\om_0 }{ 4R^6}
\left( \frac{2d^2 }{ 3\hbar\om_0} \right)^2.
\label{mod3}
\ee
We define
\be
D(R)\equiv \frac{W_{ee}(R) }{ W_L(R)} = \frac{4}{3\pi} \int_0^{\infty} dy \left( \frac{1}{y^2+1} \right)^2
\frac{1}{\epsilon^2} [n^4r^4y^4+2n^3r^3y^3+5n^2r^2y^2+6nry+3]e^{-2nry},
\label{mod4}
\ee
where $r\equiv\om_0R/c$ and  $\eps$ and $n$ are evaluated at $i\om_0y$:
\be
\eps(i\om_0y)=n^2(i\om_0y)=1+ \frac{C}{y^2+1},
\label{mod5}
\ee
where $C=8\pi Nd^2/3\hbar\om_0$. In the limit $C=0$ and $r\rightarrow 0$, $D\rightarrow 1$;
 for $r\rightarrow\infty$, $D\rightarrow 23/3\pi r$, or $W_{ee}=-23\hbar c\alpha^2(0)/4\pi R^7$, the
famous Casimir-Polder result. Figure 1 plots $D(R)$ for $C=0$ (vacuum) and $C=3$.

This simple model is not in any sense meant to be a realistic, quantitative description of the van der
Waals interaction between two atoms embedded in a dielectric. It does, however, suggest that the
predominant effect of the host medium on the van der Waals interaction is to weaken it, without
substantially changing the distance dependence in either the nonretarded or retarded regimes. More
realistic models of the van der Waals interaction in a liquid, for example, must take into account
local field corrections, as has been discussed, for instance, by Abrikosov, {\it et al.} \cite{Abrikosovbook}
and McLachlan \cite{mcl}.

\begin{figure}[t]
\includegraphics[width=8cm]{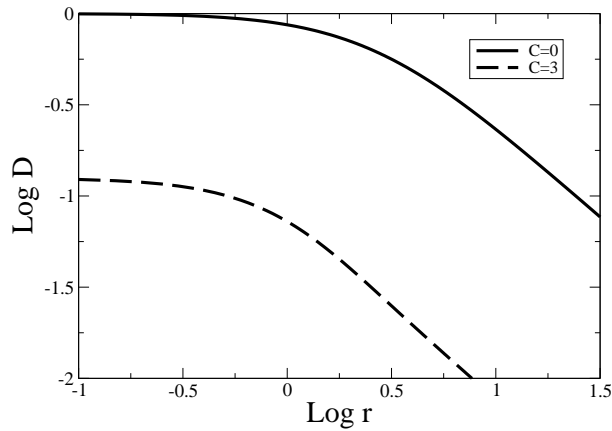}
\caption{The van der Waals interaction (\ref{result1}) divided by the London interaction for a two-level model.
The upper curve is for two atoms in vacuum ($C=0$), the lower curve for two atoms in a dielectric defined by
(\ref{mod5}) with $C=3$.}
\end{figure}

Our results for the electric-electric, magnetic-magnetic, and electric-magnetic van der Waals interactions
between two atoms embedded in a dispersive magneto-dielectric medium are in agreement with those obtained
previously \cite{TomasPRA05}-\cite{Welsch2atoms}. Unlike previous
derivations, however, we obtained the interactions directly rather than inferring them from
the dilute-medium limit of van der Waals interactions between macroscopic bodies, where the interactions can
be obtained by pairwise summations of interatomic interactions \cite{Lifshitz56}-\cite{Abrikosovbook}.

What is perhaps more interesting, however, concerns the role of absorption in the host medium. In the preceding
section we accounted for dissipation (absorption) in the host magneto-dielectric, as in previous work. In
Section \ref{sec2}, however, we ignored any possibility that the medium could be absorbing: we worked
with expressions for the electric and magnetic fields that derive directly from the assumption that the
medium is non-absorbing \cite{Milonni2003}.

The same situation holds, for instance, in the case of the Lifshitz formula for the van der Waals interaction
between infinite, plane-parallel dielectric media. Lifshitz's original derivation, and various derivations
that followed, include the imaginary (dissipative) part of the dielectric function $\epsilon(\omega)$. Derivations
of the Lifshitz formula based on changes in zero-point field energy arising from the dielectric media, however,
make no reference to dissipation \cite{Barash}, \cite{Vankampetc}, \cite{Milonnibook}. In other words, the Lifshitz formula can be derived without explicit accounting for absorption.

Ginzburg \cite{Ginzburgbook} has also noted that (macroscopic) van der Waals interactions can be correctly
derived based on changes in zero-point field energy, without accounting for absorption. He remarks, in
connection with such derivations \cite{Vankampetc}, that ``oddly enough there is no mention that they consider
directly only transparent media," and then gives reasons why the van der Waals (electric-electric) interaction for
real media can be obtained by presuming non-absorbing media: ``Firstly, the permittivities ... are
functions. Secondly, the function $\epsilon(\om)$ is always real on the imaginary axis." The results (\ref{result1}),
(\ref{result2}), and (\ref{neweq4}), for instance, all involve permittivities and permeabilities on the imaginary
axis.

A simple and more physical explanation can be given for why (ground-state) van der Waals interactions calculated for
non-absorbing media apply directly to real (absorbing) media, as we have found for the electric-electric, magnetic-magnetic,
and electric-magnetic van der Waals interactions between two atoms embedded in a magnetic-dielectric medium. At
zero-temperature, for instance, any atom of the host medium is in its ground state and can absorb
radiation that is resonant with one of its transitions to an excited state. It cannot, of course, absorb
from the vacuum field: in this case the fluctuations in the field that might induce absorption are exactly
cancelled by fluctuations in the atom itself \cite{Milonnibook}. The same is true when the atom is part
of a magneto-dielectric in which is embedded, as in examples considered in this paper, two guest atoms.
Any atom of the host medium still finds itself in a vacuum field state, regardless of the nature or
the number of guest atoms. The host and guest atoms modify the modes of the field from the simple
plane waves of a pure vacuum, but the field remains in a vacuum state $|{\rm vac}\rangle$
($f_{\lambda,i}|{\rm vac}\rangle=0$ for any mode). Just as in free space, therefore, there is no absorption
unless it is possible to populate one or more field modes; this would be the case only if there were an
applied external field or if one or more atoms is excited, leading to the possibility that a
different atom could absorb its emitted ({\it real}) photon. In other words, absorption by the
 host medium would play a role if we were to consider a
van der Waals interaction involving excited atoms. Otherwise one can expect to obtain correct van der Waals
interactions without having to account for the absorption that is always present in a real medium. This
expectation applies, of course, regardless of how many atoms are involved and regardless of the shape
of any macroscopic bodies for which the van der Waals forces are to be calculated.

In light of recent interest in negative-index media \cite{Pendryreview,Veselagoreview}, let us reconsider specifically the most important
of the van der Waals interactions we have calculated in this paper, namely that between two electrically
polarizable atoms. First we note that the expressions for the quantized fields in Section \ref{sec2} are
directly applicable to negative-index media \cite{Milonni2003}. In a negative index medium $n$, $\eps$, and $\mu$
are all negative at some frequency or range of frequencies. This would at first glance suggest that the dipole-dipole
interaction $V_{ij}^{ee}(\om,\bR)$ [Eq. (\ref{electric_dipole_tensor})] changes sign at frequencies for which the
refractive index is negative. To see that this is not the case, note that Eq. (\ref{Vee}) is unchanged whenever
$n(\om)$, $\eps(\om)$, and $\mu(\om)$ all change sign within any frequency range. This means that the dipole-dipole
interaction in a negative-index medium does not change sign, and in particular that (\ref{electric_dipole_tensor}) is directly applicable in general provided we just replace $n(\om)$, $\eps(\om)$, and $\mu(\om)$ by their absolute values.
It follows similarly that the van der Waals interaction does not change sign or undergo any other significant
change in a magneto-dielectric medium. The same conclusion applies to the magnetic-magnetic and mixed van der Waals
interactions, and is in agreement with the conclusions of Buhmann {\it et al.} \cite{WelschEPJD05}.

The pairwise electric-electric and magnetic-magnetic van der Waals interactions are always attractive, whereas
the pairwise electric-magnetic interaction is always repulsive. These results apply also in the case
of negative-index media, at least to the extent that such media can be modeled as continua. It is well known,
however, that non-pairwise van der Waals interactions can be repulsive \cite{teller}. Evidently repulsive Casimir effects such as those recently suggested by Henkel and Joulain \cite{HenkelEPL05} and by Leonhardt and Philbin \cite{ulf} in the case of negative-index media must in some way involve either non-pairwise interactions or electric-magnetic
van der Waals interactions.

%##################################################################################
%   THE BIBLIOGRAPHY
% ##################################################################################

\end{document}